\documentclass[aip,jmp]{revtex4-1}

\usepackage{amssymb}
\usepackage{latexsym}
\usepackage{amsmath,amsthm}
\usepackage{tikz}
\usepackage{hyperref}
\usepackage{graphicx}
\usepackage{color}
\usepackage{tikz}
\usetikzlibrary{decorations.markings}

\usepackage{verbatim}

\newcommand{\Su}{\sum_{b = 1}^B}
\newcommand{\cb}{\cos\theta_b}

\newtheorem{thm}{Theorem}

\newcommand{\ud}{\mathrm{d}}
\newcommand{\ui}{\mathrm{i}}
\newcommand{\UI}{\mathrm{I}}
\newcommand{\SU}{\mathrm{SU}}
\newcommand{\ue}{\mathrm{e}}
\newcommand{\nn}{\nonumber}
\newcommand{\bmu}{\mu}
\newcommand{\bhmu}{\hat{\mu}}
\newcommand{\bpsi}{\boldsymbol{\psi}}

\newcommand{\A}{\mathbb{A}}
\newcommand{\B}{\mathbb{B}}
\renewcommand\Re{\operatorname{Re}}
\renewcommand\Im{\operatorname{Im}}
\newcommand{\tk}{\tilde{k}}

\newcommand{\csch}{\mathrm{csch}}

\begin{document}

\title{Zeta Functions of the Dirac Operator on Quantum Graphs}
\date{\today}
\author{JM Harrison}
\email{jon\_harrison@baylor.edu}
\author{T Weyand}
\email{tracy\_weyand@baylor.edu}
\affiliation{Department of Mathematics, Baylor University, Waco, TX 76798, USA}
\author{K Kirsten}
\email{klaus\_kirsten@baylor.edu}
\affiliation{GCAP-CASPER, Department of Mathematics, Baylor University, Waco, TX 76798, USA}

\begin{abstract}
We construct spectral zeta functions for the Dirac operator on metric graphs. We start with the case of a rose graph, a graph with a single vertex where every edge is a loop.
The technique is then developed to cover any finite graph with general energy independent matching conditions at the vertices.
The regularized spectral determinant of the Dirac operator is also obtained as the derivative of the zeta function at a special value.
In each case the zeta function is formulated using a contour integral method, which extends results obtained for Laplace and Schr\"{o}dinger operators on graphs.
\end{abstract}

\pacs{03.65.-w, 02.10.Ox, 03.65.Pm}


\maketitle

\section{Introduction}

Quantum graphs are well studied objects in mathematical physics where they are typically employed to model properties of quasi-one-dimensional structures like carbon nanotubes and photonic crystals, or to investigate the quantum mechanics of  systems with chaotic classical dynamics\cite{BK_book, K04}.  While most investigation of quantum graphs focuses on the Laplace and Schr\"odinger operators, there has also been recent interest in properties of the Dirac operator, particularly in relation to thin carbon structures where surprisingly the Dirac equation provides the effective model for non-relativistic electronic properties of the system\cite{CGP09, HRP10, NGM05}.

In this article we construct spectral zeta functions of the Dirac operator on metric graphs.
A lot of work has gone into understanding the Ihara zeta function associated with combinatorial graphs.  The Ihara zeta function of a combinatorial graph is defined by an Euler product over the set of primitive closed loops which do not involve backtracking (primitive periodic orbits)\cite{H89, ST96, S86}.  In addition to the adjacency structure of a combinatorial graph, a quantum graph consists of intervals connecting pairs of adjacent vertices equipped with a differential operator.  Here, rather than the Ihara zeta function, it is natural to consider the zeta function associated with the point spectrum $\{\lambda_j \}$ of the quantum graph,
\begin{equation}
\zeta (s) = {\sum_{j}}^\prime \lambda_j^{-s} \ ,
\end{equation}
where the prime denotes that any eigenvalues of zero are omitted.  If the eigenvalues are the positive integers this is just the Riemann zeta function.
The results that we report here for the Dirac operator extend similar results obtained by some of the authors for the zeta functions of the Laplace\cite{HK11} and Schr\"odinger operators\cite{HKT12}.

The article is laid out as follows.  Section \ref{sec: Dirac graph} defines the Dirac operator on a graph.  In section \ref{sec: rose} we construct the spectral zeta function of the Dirac operator on a rose graph, a graph with a single vertex where every edge is a loop.  This explicit example contains the main features of the contour integral approach we adopt.   Developing this technique we formulate the zeta function of a general graph Dirac operator first without mass, section \ref{sec: general zeta no mass}, and then with mass, section \ref{sec: general zeta mass}.  As a corollary the zeta-regularized spectral determinant is obtained from the zeta function in each case.  The rose graph example we use has a natural analogy with the well studied case of the Laplacian on a star graph with the standard Neumann-like vertex conditions, where the wavefunction is continuous at vertices and the outgoing derivatives sum to zero.  As the Dirac extension of this canonical example is not well known, for completeness, we include the derivation of the secular equation of the rose graph as an appendix.

\section{The Dirac operator on a graph}\label{sec: Dirac graph}

\begin{figure}[tbh]
\begin{center}
\begin{tikzpicture}
  [scale=1.5,every node/.style={circle,fill=black!, scale=.5}]
  \node (n1) at (-1,0) {};
  \node (n2) at (.3,1)  {};
  \node (n3) at (1,0)  {};
  \node (n4) at (0,-.7) {};
  \node (n5) at (-2,0) {};
  \node (n6) at (-.2,0)  {};

  \foreach \from/\to in {n1/n5,n1/n2,n1/n4,n2/n3,n3/n4, n1/n6, n6/n3, n6/n2}
    \draw (\from) -- (\to);

\end{tikzpicture}
\end{center}
\caption{A graph with $6$ vertices and $8$ bonds.}\label{fig: general graph}
\end{figure}
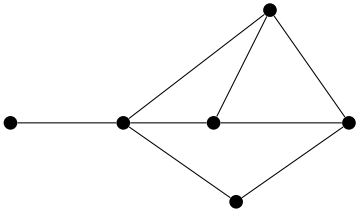

A graph $G$ consists of a set  of vertices $\mathcal{V}$ with some pairs of vertices connected by bonds, so a bond $b=(u,v)$ consists of an unordered pair of vertices $u,v \in \mathcal{V}$; see figure \ref{fig: general graph}.  We extend $G$ to a metric graph where each bond $b=(u,v)$ is associated with an interval $[0,L_b]$.  The vertices $u$ and $v$ lie at the ends of the interval and the choice of orientation for the interval will turn out to be arbitrary.  We refer to $L_b$ as the length of $b$, and the bonds are enumerated so that $b \in \mathcal{B}=\{1, 2, \ldots , B\}$. Here we consider only finite graphs where there are a finite number of bonds $B$ and the length of every interval is finite.

A \emph{quantum graph} is a metric graph together with a differential operator that acts on functions defined on the set of intervals associated with the bonds.
In this paper, we consider the one-dimensional time-independent Dirac operator on the intervals,
\begin{equation}\label{eq: Dirac operator}
\mathcal{D} := -\ui\hbar c\alpha \frac{\ud}{\ud x} + mc^2\beta \ ,
\end{equation}
where $\alpha$ and $\beta$ are matrices that satisfy $\alpha^2 = \beta^2 = \UI$ and $\alpha\beta + \beta\alpha = 0$, the Dirac algebra in one dimension.  To simplify notation, from now on we assume $\hbar=c=1$.
We may think of a Dirac operator on a metric graph as representing the restriction of the Dirac equation in three dimensions to a one-dimensional network.  Hence, it is natural to require $\alpha$ and $\beta$ to be $4\times 4$ matrices. (One might instead choose $\alpha$ and $\beta$ to be $2\times 2$ matrices, the simplest irreducible representation of the one-dimensional Dirac algebra.  However, in this case the wavefunctions depend of the orientation of the intervals which is unphysical.  To resolve this and allow time-reversal invariance, one must use pairs of edges, one oriented in each direction\cite{BH03}.)
The Hilbert space for our operator is the direct sum of Hilbert spaces for each bond
\begin{equation}\label{eq: Dirac domain}
\mathcal{H} = \bigoplus_{b=1}^B L^2([0,L_b]) \otimes \mathbb{C}^4.
\end{equation}

To fix a domain of functions in $\mathcal{H}$ on which $\mathcal{D}$ is self-adjoint, we must introduce appropriate matching conditions at the vertices.
General vertex conditions can be specified via a pair of $4B\times 4B$ matrices $\A$ and $\B$.  A function $\psi \in \mathcal{H}$ satisfies the boundary conditions if
\begin{equation}\label{eq: self-adjoint condition}
\A\bpsi^+ + \B\bpsi^- = 0
\end{equation}
where
\begin{align}\label{eq:defn psi+-}
    \bpsi^+ &= \Big( \psi_1^1(0),\psi_2^1(0),\dots,\psi_1^B(0),\psi_2^B(0),
    \psi_1^1(L_1),\psi_2^1(L_1),\dots,\psi_1^B(L_B),\psi_2^B(L_B) \Big)^T  \ ,\\
    \bpsi^- &= \Big( -\psi_4^1(0),\psi_3^1(0),\dots,-\psi_4^B(0),\psi_3^B(0),
    \psi_4^1(L_1),-\psi_3^1(L_1),\dots,\psi_4^B(L_B),-\psi_3^B(L_B) \Big)^T \ .
\end{align}
Here $\psi_j^b(x_b)$ is the $j$-th component of the four component function on the bond $b$.  Then $\mathcal{D}$ is self-adjoint if and only if
\begin{equation*}
    \textrm{rank}(\A,\B)=4B \qquad \textrm{and} \qquad \A\B^\dag=\B\A^\dag \ .
\end{equation*}
This classification of self-adjoint vertex matching conditions was obtained by Bolte and Harrison\cite{BH03} and emulates a typical classification of self-adjoint Laplacians on graphs by Kostrykin and Schrader \cite{KS99}.

Eigenspinors $\psi_k$ that satisfy $\mathcal{D}\psi_k = E(k)\psi_k$ are comprised of plane waves on each bond.
While $\alpha$ and $\beta$ are only required to satisfy the Dirac algebra, in order to make the  calculations explicit, we choose
\begin{equation}\label{eq: alpha and beta}
\alpha = \begin{pmatrix} 0 & 0 & 0 & -\ui\\0 & 0 & \ui & 0\\0 & -\ui & 0 & 0\\\ui & 0 & 0 & 0\end{pmatrix}\ ,
\hspace{1cm}
\beta = \begin{pmatrix}1 & 0 & 0 & 0\\ 0 & 1 & 0 & 0\\ 0 & 0 & -1 & 0\\ 0 & 0 & 0 & -1\end{pmatrix} \ .
\end{equation}
Then, for positive energy, the eigenspinors have the form
\begin{align}\label{eq: eigenspinors}
\psi^b(x_b) = \mu^b_\alpha &
\begin{pmatrix}1\\0\\0\\\ui\gamma(k)\end{pmatrix}\ue^{\ui kx_b} + \mu^b_\beta\begin{pmatrix}0\\1\\-\ui\gamma(k)\\0\end{pmatrix}\ue^{\ui kx_b}
+
\hat{\mu}^b_\alpha\begin{pmatrix}1\\0\\0\\-\ui\gamma(k)\end{pmatrix}\ue^{-\ui kx_b} + \hat{\mu}^b_\beta\begin{pmatrix}0\\1\\\ui\gamma(k)\\0\end{pmatrix}\ue^{-\ui kx_b}
\end{align}
with
\begin{equation}\label{eq: gamma k and E}
\gamma(k) := \frac{\sqrt{k^2+m^2} - m}{k} \ ,\hspace{1cm} E(k) = \sqrt{k^2 + m^2} \ ,
\end{equation}
while negative energy eigenspinors that satisfy $\mathcal{D}\psi_k = -E(k)\psi_k$ have the form,
\begin{align}\label{eq: neg eigenspinors}
\psi^b(x_b) = \mu^b_\alpha &
\begin{pmatrix}\ui\gamma(k)\\0\\0\\1\end{pmatrix}\ue^{\ui kx_b} + \mu^b_\beta\begin{pmatrix}0\\-\ui\gamma(k)\\1\\0\end{pmatrix}\ue^{\ui kx_b}
+
\hat{\mu}^b_\alpha\begin{pmatrix}-\ui\gamma(k)\\0\\0\\1\end{pmatrix}\ue^{-\ui kx_b} + \hat{\mu}^b_\beta\begin{pmatrix}0\\\ui\gamma(k)\\1\\0\end{pmatrix}\ue^{-\ui kx_b}
\end{align}
 with the same definitions of $\gamma$ and $E$.

Taking the positive energy case, the matching conditions \eqref{eq: self-adjoint condition} take the form,
\begin{equation}\label{eq: boundary conditions as matrices}
\left(\A\begin{pmatrix}
\UI_{2B} & \UI_{2B}\\
\ue^{\ui kL} & \ue^{-\ui kL}
\end{pmatrix}
+ \ui\gamma(k)\B\begin{pmatrix}
-\UI_{2B} & \UI_{2B}\\
\ue^{\ui kL} & -\ue^{-\ui kL}\\
\end{pmatrix}
\right)\begin{pmatrix}
\mu\\\hat{\mu}
\end{pmatrix}=0 \ ,
\end{equation}
where
$\ue^{\ui k L}$ denotes the diagonal matrix $\textrm{diag}\{\ue^{\ui k L_1},\ue^{\ui k L_1},\dots,\ue^{\ui k L_B},\ue^{\ui k L_B}\}$ and furthermore
$L =\textrm{diag}\{ L_1,L_1,L_2,L_2,\ldots,L_B,L_B \}$ is used to define other diagonal matrices similarly.  The vector
$\mu = (\mu_\alpha^1,\mu_\beta^1,\mu_\alpha^2,\mu_\beta^2,\ldots,\mu_\alpha^B,\mu_\beta^B)^T$ and the vector $\hat{\mu}$ is defined similarly.
We see that if the coefficients $\mu$ and $\hat{\mu}$ define an eigenspinor, then,
\begin{equation}\label{eq: determinant}
\det\left(\A\begin{pmatrix}
\UI_{2B} & \UI_{2B}\\
\ue^{\ui kL} & \ue^{-\ui kL}
\end{pmatrix}
+ \ui\gamma(k)\B\begin{pmatrix}
-\UI_{2B} & \UI_{2B}\\
\ue^{\ui kL} & -\ue^{-\ui kL}\\
\end{pmatrix}
\right) = 0.
\end{equation}
For $k\notin \{ n\pi/L_b \}_{n\in \mathbb{N}, b\in \mathcal{B}}$, multiplying on the right by
\begin{equation}\label{eq: inverse matrix}
\det\left(\begin{pmatrix} \UI_{2B} & I_{2B}\\ \ue^{\ui kL} & \ue^{-\ui kL}\end{pmatrix}^{-1}\right)
= \det\left( \begin{pmatrix} \ue^{-\ui kL} & - \UI_{2B}\\-\ue^{\ui kL} & \UI_{2B}\end{pmatrix}  \begin{pmatrix} \frac{-1}{2\ui \sin kL} &0 \\  0 & \frac{-1}{2\ui \sin kL}  \end{pmatrix}\right),
\end{equation}
we obtain instead of (\ref{eq: determinant}),
\begin{equation}\label{eq: general secular eqn}
\det\left(\A + \gamma(k)\B\begin{pmatrix}
\cot kL & -\csc kL\\
-\csc kL & \cot kL
\end{pmatrix}\right) = 0 \ .
\end{equation}
 Equation \eqref{eq: general secular eqn} is a \emph{secular equation} for the Dirac operator.  Roots $k_j$ of the left hand side correspond to energy eigenvalues according to \eqref{eq: gamma k and E}.  Notice that in this form the matching conditions appear explicitly in the secular equation.

Following the same argument for negative energy eigenspinors one obtains the secular equation,
\begin{equation}\label{eq: general secular eqn -ve E}
\det\left(\gamma(k) \A - \B\begin{pmatrix}
\cot kL & -\csc kL\\
-\csc kL & \cot kL
\end{pmatrix}\right) = 0 \ .
\end{equation}
We will denote roots of the negative energy secular equation $\tk_j$, where each root corresponds to a negative eigenvalue $-\sqrt{\tk_j^2+m^2}$.

If $m=0$ the positive energy secular equation \eqref{eq: general secular eqn} reads,
\begin{equation}\label{eq: general secular eqn m=0}
\det\left(\A + \B\begin{pmatrix}
\cot kL & -\csc kL\\
-\csc kL & \cot kL
\end{pmatrix}\right) = 0 \ ,
\end{equation}
and the negative energy equation \eqref{eq: general secular eqn -ve E} reads,
\begin{equation}\label{eq: general secular eqn m=0 -ve E}
\det\left(\A - \B\begin{pmatrix}
\cot kL & -\csc kL\\
-\csc kL & \cot kL
\end{pmatrix}\right) = 0 \ .
\end{equation}
Changing $k$ to $-k$ we see that the equations agree.  This is not surprising as, with $m=0$, changing the sign of $k$ changes positive to negative energy eigenspinors and vice versa.
Hence for $m=0$ we consider \eqref{eq: general secular eqn m=0} to be the single secular equation whose positive roots are the positive energy eigenvalues and whose negative roots are the negative energy eigenvalues.

For matrices $\A$ and $\B$ that define a self-adjoint realization of the Dirac operator, $(\A - \ui\gamma(k)\B)$ is invertible and we can write \eqref{eq: boundary conditions as matrices} in the form,
\begin{equation}\label{eq: matrix conditions}
\tilde{\mu} =- (\A - \ui\gamma(k)\B)^{-1}(\A + \ui\gamma(k)\B)\begin{pmatrix}0&\ue^{\ui kL}\\\ue^{\ui kL}&0\end{pmatrix}\tilde{\mu},
\end{equation}
where
\begin{equation}
\tilde{\mu} = \begin{pmatrix}\UI & 0\\0&\ue^{-\ui kL}\end{pmatrix}\begin{pmatrix}\mu\\\hat{\mu}\end{pmatrix} \ .
\end{equation}
Then
\begin{equation}\label{eq: transition matrix}
T = -(\A - \ui\gamma(k)\B)^{-1}(\A + \ui\gamma(k)\B)
\end{equation}
is called the  \textit{transition matrix}.
When $\A$ and $\B$ define a self-adjoint realization of the Dirac operator, $T$ is unitary \cite{BH03}.  Applying $T$ to a vector of incoming plane wave coefficients at the vertices of the graph yields a vector of outgoing coefficients at the vertices.  The matrix $\begin{pmatrix}0 & \ue^{\ui kL}\\\ue^{\ui kL}& 0\end{pmatrix}$ in \eqref{eq: matrix conditions} transforms outgoing coefficients to incoming coefficients at the opposite end of the bond; the phase of the coefficient at the two ends of a bond $b$ differs by $\ue^{\ui k L_b}$.  Alternatively the transition matrix can be obtained by combining scattering matrices at each of the vertices.
From \eqref{eq: matrix conditions} we see that an eigenspinor is defined by a vector $\tilde{\mu}$ of plane wave coefficients invariant under the action of
\begin{equation*}
T\begin{pmatrix}0 & \ue^{\ui kL}\\\ue^{\ui kL}& 0\end{pmatrix} \ ,
\end{equation*}
which is referred to as the \textit{quantum evolution operator} \cite{BK_book}.
For such an eigenspinor,
\begin{equation}
\det\left(\UI - T\begin{pmatrix}0 & \ue^{\ui kL}\\\ue^{\ui kL}& 0\end{pmatrix}\right) = 0 \ ,
\end{equation}
 an alternate form of the secular equation.

\subsection{Effect of time-reversal symmetry}\label{sec: time reversal}

For the Dirac equation the standard representation of the time-reversal operator is,
\begin{equation}\label{eq: time-reversal op}
    \mathcal{T}=\ui \begin{pmatrix}
J&0\\
0&J\\
\end{pmatrix} \mathcal{K}\ ,
\end{equation}
where $\mathcal{K}$ is complex conjugation and $
J = \begin{pmatrix}
0&1\\-1&0
\end{pmatrix}$.  $\mathcal{T}$ is an anti-unitary operator which squares to $-\UI$.
As shown in \cite{BH03}, time-reversal symmetry requires the transition matrix satisfy,
\begin{equation}\label{eq: time-reversal symm Dirac}
T^T = \begin{pmatrix}
J^{-1}\\&\ddots\\&  & J^{-1}
\end{pmatrix}T\begin{pmatrix}
J\\&\ddots\\&&J
\end{pmatrix} \ .
\end{equation}
If we choose matrices $\A$ and $\B$ of the form
\begin{equation}\label{eq: AB time rev symm}
\A = \left(\widetilde{\A} \otimes \UI_2\right)U \hspace{1cm} \B = \left(\widetilde{\B}\otimes \UI_2\right)U
\end{equation}
where $U$ is a block diagonal matrix
\begin{equation}\label{eq: defn of U}
    U= \textrm{diag} \{ u^1_o , \dots, u^B_o ,u^1_t, \dots, u^B_t \}
\end{equation}
with $u^b \in \mathrm{SU}(2)$ and $o,t$ standing for the origin and terminus of the bond $b$ respectively, then the Dirac operator is self-adjoint if $\mathrm{rank}(\widetilde{\A},\widetilde{\B})=2B$ and $\widetilde{\A}\widetilde{\B}^\dag$ is self-adjoint.  Furthermore, if
\begin{equation}\label{eq: transition matrix}
\widetilde{T} := -(\widetilde{\A} - \ui \widetilde{\B})^{-1}(\widetilde{\A} + \ui \widetilde{\B})
\end{equation}
is symmetric, then $T$ satisfies \eqref{eq: time-reversal symm Dirac} and $\A$ and $\B$ define time-reversal symmetric boundary conditions for the Dirac operator.  Time-reversal symmetry for the Laplace operator requires a symmetric transition matrix, so matching conditions $\widetilde{\A}$ and $\widetilde{\B}$ that define self-adjoint time-reversal symmetric matching conditions for the Laplacian provide a pair of matrices $\A$ and $\B$ defining a self-adjoint time-reversal symmetric realization of the Dirac operator \cite{BH03}.

If we consider the case of a graph with time-reversal symmetry where the vertex matching conditions have the form \eqref{eq: AB time rev symm}, the secular equation \eqref{eq: general secular eqn} simplifies significantly,
\begin{equation}\label{eq: secular eqn simplify 1}
\det\left(\left(\widetilde{\A} \otimes \UI_2\right)U + \gamma(k)
\left(\widetilde{\B}\otimes \UI_2\right)U
\begin{pmatrix}
\cot k L  & -\csc k L  \\
-\csc k L & \cot k L
\end{pmatrix}\right) = 0 \ .
\end{equation}
Post multiplying by $U^{-1}$,
\begin{equation}\label{eq: secular eqn simplify 1}
\det\left(\left(\widetilde{\A} \otimes \UI_2\right) + \gamma(k)
\left(\widetilde{\B}\otimes \UI_2\right)
\begin{pmatrix}
\cot k L  & -\csc k L W \\
-\csc k L  W^{-1} & \cot kL
\end{pmatrix}\right) = 0 \ ,
\end{equation}
where $W=\textrm{diag} \{ w^1 , \dots, w^B \}$ with $w^b=u^b_o (u^b_t)^{-1} \in \mathrm{SU}(2)$.  Each $w^b$ can be diagonalized so
\begin{equation}\label{eq:diag w}
    w^b = (\omega^b)^{-1}
    \begin{pmatrix}
    \ue^{\ui \theta_b} & 0 \\
    0 & \ue^{-\ui \theta_b}
\end{pmatrix} \omega^b \ .
\end{equation}
Setting $\Omega=\textrm{diag}\{ \omega^1,\dots \omega^B \}$ and noting that $\Omega$ commutes with  $\csc k L $ and $\cot k L $,
\begin{align}
    &\left(\left(\widetilde{\A} \otimes \UI_2 \right) + \gamma(k)
\left(\widetilde{\B}\otimes \UI_2\right)
\begin{pmatrix}
\cot k L  & -\csc k L W \\
-\csc k L  W^{-1} & \cot k L
\end{pmatrix} \right)= \nonumber \\
 \begin{pmatrix}
\Omega^{-1} & 0\\
0 &\Omega^{-1}
\end{pmatrix}
&\left(
\left(\widetilde{\A} \otimes \UI_2 \right) + \gamma(k)
\left(\widetilde{\B}\otimes \UI_2\right)
\begin{pmatrix}
\cot k L  & -\csc k L D \\
-\csc k L  D^{-1} & \cot k L
\end{pmatrix}\right)
\begin{pmatrix}
\Omega & 0\\
0 &\Omega
\end{pmatrix} \ ,
\end{align}
where
$D=\textrm{diag}\{\ue^{\ui \theta_1}, \ue^{-\ui \theta_1},\ue^{\ui \theta_2},\dots,\ue^{-\ui \theta_B} \}$.  Since $\det \Omega =1$,
\begin{equation}\label{eq: secular eqn simplify 2}
    \det \left(
\left(\widetilde{\A} \otimes \UI_2\right) + \gamma(k)
\left(\widetilde{\B}\otimes \UI_2\right)
\begin{pmatrix}
\cot k L  & -\csc k L D \\
-\csc k L  D^{-1} & \cot k L
\end{pmatrix}\right) = 0 \ .
\end{equation}
Now, applying row and column permutations to separate the odd and even rows and columns, we see that
\begin{align}
   & \det \left(
\widetilde{\A}  + \gamma(k)\widetilde{\B}
\begin{pmatrix}
\cot k \widetilde{L} & -\csc k \widetilde{L} \, \ue^{\ui \boldsymbol{\theta}} \\
-\csc k \widetilde{L} \, \ue^{-\ui \boldsymbol{\theta}} & \cot k\widetilde{L}
\end{pmatrix}\right)  \nn \\
& \qquad \times
 \det
    \left(
\widetilde{\A}  + \gamma(k)\widetilde{\B}
\begin{pmatrix}
\cot k\widetilde{L} & -\csc k\widetilde{L} \, \ue^{-\ui \boldsymbol{\theta}} \\
-\csc k\widetilde{L} \, \ue^{\ui \boldsymbol{\theta}} & \cot k\widetilde{L}
\end{pmatrix}\right)
=0 \ ,
\end{align}
where $\widetilde{L}=\textrm{diag}\{ L_1,\dots,L_B \}$.  Notice that if the matrices $\A$ and $\B$ defining the vertex conditions are real this is just,
 \begin{equation}
   \left| \det \left(
\widetilde{\A}  + \gamma(k)\widetilde{\B}
\begin{pmatrix}
\cot k \widetilde{L} & -\csc k \widetilde{L} \, \ue^{\ui \boldsymbol{\theta}} \\
-\csc k \widetilde{L} \, \ue^{-\ui \boldsymbol{\theta}} & \cot k\widetilde{L}
\end{pmatrix}\right) \right|^2
=0 \ .
\end{equation}
In either case the reduced form of the secular equation allow the subsequent results for the zeta function to be evaluated more easily in the presence of time-reversal invariance for any particular graph.
%


\subsection{Spectral zeta function}

In this paper we construct and analyze the spectral zeta function of the Dirac operator on a quantum graph.
The spectral zeta function generalizes the Riemann zeta function where the sum over integers is replaced with a sum over the graph eigenvalues,
\begin{align}\label{eq: general zeta function}
\zeta(s) &=  2 \sum_{j=1}^{\infty}\,\!^{^\prime} \left[ E(k_j)^{-s} + \big(-E(\tk_j)\big)^{-s} \right] \nn \\
&= 2 \sum_{j=1}^\infty\,\!^{^\prime} \left[(k_j^2 + m^2)^{-s/2} +(-1)^{-s} (\tk_j^2 + m^2)^{-s/2}\right] \ .
\end{align}
The factor of two simply incorporates  Kramers' degeneracy and the prime indicates that any zero eigenvalues are ignored.  The sum runs over positive roots of both the positive and negative energy secular equations.  In the zero mass case this simplifies, and
\begin{equation}\label{eq: general zeta function m=0}
\zeta(s)= 2 \sum_{j=-\infty}^{\infty}\,\!\!\!^{^\prime} \,\, k_j^{-s} \ ,
\end{equation}
where $k_j$ is a positive or negative non-zero root of (\ref{eq: general secular eqn m=0}).

\section{The Dirac rose graph}\label{sec: rose}

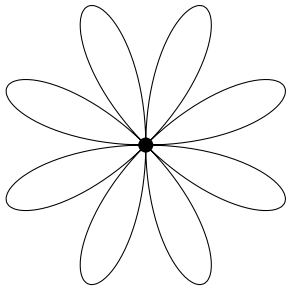
\begin{figure}[tbhp!]
\begin{center}
\begin{tikzpicture}
\draw[domain=0:6.28,samples=200,smooth] plot (xy polar
cs:angle=\x r,radius=      {2*sin(4*\x r)});    
\fill[black] (0,0) circle (.1cm);
\end{tikzpicture}
\end{center}
\caption{\label{fig: rose graph} A rose graph with eight bonds.}
\end{figure}

We will construct the zeta functions of general graph Dirac operators.  However, to illustrate the technique, we first consider an example which can be analyzed more explicitly, the rose graph.  We will see that this is the Dirac analogue of the star graph with Neumann-like matching conditions that is frequently studied in the case of the Laplace operator.
A rose graph consists of one vertex and $B$ bonds where each bond is a loop that begins and ends at the vertex; see figure \ref{fig: rose graph}.

Let
\begin{equation}\label{eq: definition of v and w}
v^b(x_b) = \begin{pmatrix} \psi^b_1(x_b)\\\psi^b_2(x_b)\end{pmatrix} \hspace{1cm}\mbox{and}\hspace{1cm} w^b(x_b) = \begin{pmatrix}-\psi^b_4(x_b)\\\psi_3(x_b)\end{pmatrix}.
\end{equation}
We consider the following self-adjoint matching conditions at the central vertex.  Firstly
\begin{equation}\label{eq: boundary condition 1}
u^b_o v^b(0) = u^b_t v^b(L_b) = \xi
\end{equation}
for all bonds $b$ where $\xi$ is not a constant vector but rather a placeholder for the value of the $2B$ vectors which agree at the vertex.  Secondly
\begin{equation}\label{eq: boundary condition 2}
\Su u^b_o w^b(0) - \Su u^b_t w^b(L_b) = 0 \ .
\end{equation}
Recall, $u^b \in \mathrm{SU}(2)$ with $o,t$ standing for the origin and terminus of $b$.

Such matching conditions are analogous to the commonly studied Neumann-like boundary conditions for the graph Laplacian. In the Dirac case the elements of $\mathrm{SU}(2)$ generate spin rotation at the vertices.
 These vertex conditions can be encoded by a pair of matrices $\A$ and $\B$ following \eqref{eq: AB time rev symm} where $\widetilde{\A}$ and $\widetilde{\B}$ are $2B \times 2B$ matrices
\begin{equation}
\widetilde{\A} = \begin{pmatrix}
1 & -1 & 0 & \ldots & 0\\
0 & 1 & -1 & \ldots & 0\\
&& \ddots &\ddots & \\
0 & \ldots & 0 & 1 & -1\\
0 & \ldots & 0 & 0 & 0\\
\end{pmatrix} \ , \qquad
\widetilde{\B} = \begin{pmatrix}
0 & 0&\ldots & 0 \\
\vdots &\vdots&  & \vdots \\
0 & 0 & \ldots & 0 \\
1 & 1 & \ldots & 1 \\
\end{pmatrix} \ .
\end{equation}

In the case of the rose graph without mass, the secular equation \eqref{eq: general secular eqn m=0} takes a particularly simple form \cite{HarWin_JPA12},
\begin{equation}\label{eq: secular rose}
\Su \frac{\cb - \cos kL_b}{\sin kL_b} = 0,  \hspace{1cm}\mbox{where  } \cos \theta_b = \frac{1}{2}\mbox{tr}(u_o^b(u_t^b)^{-1}) \ .
\end{equation}
The argument is given in Appendix~\ref{sec: secular equation derivation} for completeness.
The roots of \eqref{eq: secular rose} are energy eigenvalues of the Dirac rose.
To see the analogy with the Laplace operator on a star graph with Neumann-like matching conditions\cite{BerBogKea_JPA01} let $\theta_b = 0$ for all bonds $b$, which corresponds to switching off the spin rotation at the vertices.  Then the secular equation becomes
\begin{equation}
\Su \tan \frac{kL_b}{2} = 0 \ ,
\end{equation}
 which is the secular equation of the Laplace operator on a star graph with Neumann-like matching conditions and bonds of length $L_b/2$.

Let
 \begin{equation}\label{eq: function f}
f(z) = z\Su \frac{\cb - \cos L_bz}{\sin L_bz}
\end{equation}
where $z \in \mathbb{C}$.
The zeros $k_j$ of $f$ are the eigenvalues of the Dirac operator while
the factor of $z$ in the definition of $f$ removes the pole at the origin.
By the argument principle, we can express the zeta function of the Dirac operator with zero mass as
\begin{equation}
\zeta(s) = \frac{1}{\pi \ui}\int _C z^{-s}\frac{\ud}{\ud z}\log f(z)\, \ud z \ ,
\end{equation}
where the contour $C$ encloses both positive and negative zeros of $f$ and avoids poles.

\begin{figure}[tbhp!]

\begin{center}
\begin{tikzpicture}
  \draw[->] (-2.4,0) -- (2.4,0);
  \draw[->] (0,-2.4) -- (0,2.4);
  \draw (-2,0) circle (.1cm);
  \draw (-1,0) circle (.1cm);
  \draw (1,0) circle (.1cm);
  \draw (2,0) circle (.1cm);
  \fill[black!40!white] (-1.5,0) circle (.1cm);
  \fill[black!40!white] (-.5,0) circle (.1cm);
  \fill[black!40!white] (.5,0) circle (.1cm);
  \fill[black!40!white] (1.5,0) circle (.1cm);
  \draw[<-,blue] (-2.5,.3) cos (-2.25,0);
  \draw[blue] (-2.25,0) sin (-2,-.3);
  \draw[blue] (-2,-.3) cos (-1.75,0);
  \draw[blue] (-1.75,0) sin (-1.5,.3);
  \draw[blue] (-1.5,.3) cos (-1.25,0);
  \draw[blue] (-1.25,0) sin (-1,-.3);
  \draw[blue] (-1,-.3) cos (-.75,0);
  \draw[blue] (-.75,0) sin (-.5,.3);
  \draw[blue] (-.5,.3) cos (-.25,0);
  \draw[blue] (-.25,0) sin (0,-.3);
  \draw[blue] (0,-.3) cos (.25,0);
  \draw[blue] (.25,0) sin (.5,.3);
  \draw[blue] (.5,.3) cos (.75,0);
  \draw[blue] (.75,0) sin (1,-.3);
  \draw[blue] (1,-.3) cos (1.25,0);
  \draw[blue] (1.25,0) sin (1.5,.3);
  \draw[blue] (1.5,.3) cos (1.75,0);
  \draw[blue] (1.75,0) sin (2,-.3);
  \draw[blue] (2,-.3) cos (2.25,0);
  \draw[blue] (2.25,0) sin (2.5,.3);
  \draw[->, blue](-2.4,-.5) -- (2.4,-.5);
  \draw[very thick] (-2.4,2.4) -- (0,0);
  \draw[red,thick,domain=0:138] plot ({cos(\x)}, {sin(\x)});
  \node[label={\small $\alpha$}] at (.5,.3){};
  \node[label={\small Contour $C$}] (a) at (1,-2){};
  \node[label={\small i)}] (a) at (-2.6,2.4){};
\end{tikzpicture}
\hspace{1cm}
\begin{tikzpicture}[decoration={markings,
mark=at position -0.3cm with {\arrow[line width=1pt]{<}},
}
]
  \draw[->] (-2.4,0) -- (2.4,0);
  \draw[->] (0,-2.4) -- (0,2.4);
  \draw (-2,0) circle (.1cm);
  \draw (-1,0) circle (.1cm);
  \draw (1,0) circle (.1cm);
  \draw (2,0) circle (.1cm);
  \fill[black!40!white] (-1.5,0) circle (.1cm);
  \fill[black!40!white] (-.5,0) circle (.1cm);
  \fill[black!40!white] (.5,0) circle (.1cm);
  \fill[black!40!white] (1.5,0) circle (.1cm);
  \draw[blue,postaction=decorate] (1,0) circle (.2cm);
  \draw[blue,postaction=decorate] (2,0) circle (.2cm);
  \draw[blue,postaction=decorate] (-1,0) circle (.2cm);
  \draw[blue,postaction=decorate] (-2,0) circle (.2cm);
  \draw[->, blue](-2.4,-2) -- (2.4,-2);
  \draw[->,blue] (-2.3,2.4) -- (0,.1);
  \draw[->,blue] (-.1,0) -- (-2.4,2.3);
  \draw[blue] (-.1,0) sin (0,-.1);
  \draw[blue] (0,-.1) cos (.1,0);
  \draw[blue] (.1,0) sin (0,.1);
  \draw[red,thick,domain=0:138] plot ({cos(\x)}, {sin(\x)});
  \draw[very thick] (-2.4,2.4) -- (0,0);
  \node[label={\small $\alpha$}] at (.5,.3){};
  \node[label={\small Contour $C'$}] at (1,-2){};
  \node[label={\small ii)}] (a) at (-2.6,2.4){};
\end{tikzpicture}
\end{center}

\caption{\label{fig:rose contour}Contours $C$ and $C'$ are shown in (i) and (ii) respectively. The branch cut of the logarithm is located at an angle $\alpha$ to the positive real axis to avoid the zeros and poles of $f$ which are shown with filled and empty circles respectively.}
\end{figure}
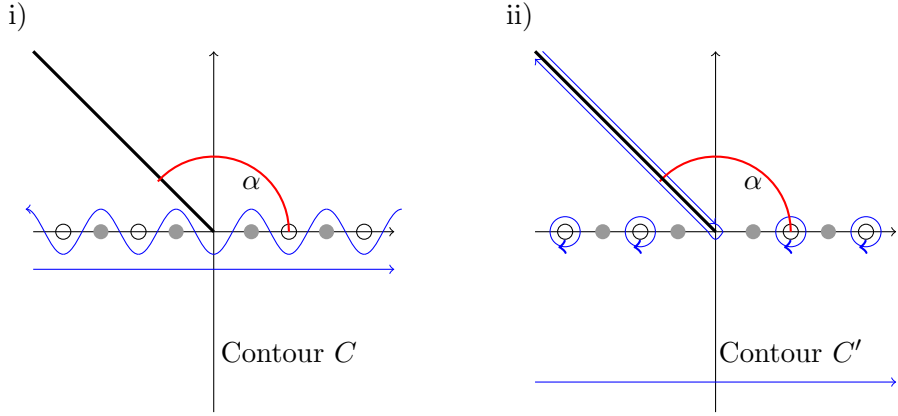

Poles of $f$ lie on the real axis at integer multiples of $\pi/L_b$ for each bond $b$.  From the structure of $f$, we can see that on the positive real axis $f$ is strictly increasing between the poles while on the negative real axis it is strictly decreasing between the poles.  Hence, for incommensurate bond lengths, $f$ has exactly one zero between each pair of adjacent poles (treating zero as a pole).  We locate the branch cut of the logarithm at an angle $\alpha$ to the positive real axis with $0<\alpha <\pi $ to avoid the zeros and poles of $f$.  Figure \ref{fig:rose contour} (i) shows the contour $C$ used to construct the zeta function.

To analyze $\zeta(s)$, we deform the contour from $C$ to $C'$, see figure \ref{fig:rose contour} (ii).  The horizontal part of the contour $C'$ will be sent to $-\infty$ in the imaginary coordinate.  Given the form of $C'$, it is natural to break $\zeta$ into three parts,
\begin{equation}\label{eq: zeta as three parts}
\zeta(s) = \zeta_p(s) + \zeta_b(s) + \zeta_l(s)\ ,
\end{equation}
where $\zeta_p$ is the contribution from the poles of $f$, $\zeta_b$ is the integral around the branch cut and $\zeta_l$ is the integral along the horizontal line.

To evaluate $\zeta_p(s)$ we must subtract residues $-z^{-s}$ at the poles $z$ of $f$,
\begin{align}
\zeta_p(s) &= 2\Su\left(\sum_{n = -\infty}^{-1} \left(\frac{n\pi}{L_b}\right)^{-s} + \sum_{n=1}^\infty \left(\frac{n\pi}{L_b}\right)^{-s}\right)\nonumber\\
&= 2(\ue^{-\ui\pi s} + 1)\zeta_R(s)\Su\left(\frac{\pi}{L_b}\right)^{-s}\ , \label{eq: zp}
\end{align}
where $\zeta_R(s)$ is the Riemann zeta function.

The integral along the horizontal line where $z=k+\ui t$ is
\begin{equation}
\zeta_l(s) =\frac{1}{\pi \ui} \int_{-\infty}^\infty (k+\ui t)^{-s} \frac{\ud}{\ud k}\log f(k+\ui t)\, \ud k \ ,
\end{equation}
which we consider in the limit $t\to -\infty$.
We observe that
\begin{align}
f(k+\ui t) &= \ui(k+\ui t)\Su \left(\frac{2\cb - (\ue^{\ui(k+\ui t)L_b}
+ \ue^{-\ui(k+\ui t)L_b})}{\ue^{\ui(k+\ui t)L_b} - \ue^{-\ui(k+\ui t)L_b}}\right) \nn \\
&\sim k+\ui t
\end{align}
and therefore $\frac{\ud}{\ud z }\log f(z) \sim 1/z$. Hence $\zeta_l(s)\to 0$
provided $\Re(s) > 0$.

Finally, the integral around the branch cut is,
\begin{align}
\zeta_b(s) &= \frac{1}{\pi \ui} \int_\infty^0 (u\ue^{\ui\alpha})^{-s} \frac{\ud}{\ud u}\log f(u\ue^{\ui\alpha})\, \ud u + \frac{1}{\pi \ui}\int_0^\infty (ue^{\ui(\alpha-2\pi)})^{-s}\frac{\ud}{\ud u} \log f(u\ue^{\ui(\alpha-2\pi)})\, \ud u \\
&= \ue^{\ui(\pi-\alpha)s}\frac{2\sin \pi s}{\pi}\int_0^\infty u^{-s}\frac{\ud}{\ud u} \log f(u\ue^{\ui\alpha})\, \ud u \ . \label{eq: zb}
\end{align}
Repeating the argument used for $\zeta_l$, we see $\frac{\ud}{\ud u} \log f(u\ue^{\ui\alpha}) \sim 1/u$ as $u \to \infty$.   Hence the integral \eqref{eq: zb} converges at infinity provided $\Re(s) > 0$. On the other hand,
\begin{equation}
f(0) = \Su \frac{\cb -1}{L_b} \ ,
\end{equation}
and as $u \to 0$
\begin{align}
\frac{\ud}{\ud u} f(u\ue^{\ui\alpha}) &= \ue^{\ui\alpha}\Su \frac{\cb - \cos uL_b \ue^{\ui\alpha}}{\sin uL_b \ue^{\ui\alpha}} + u \ue^{\ui\alpha}\Su \frac{L_b \ue^{\ui\alpha}(1 - \cb \cos uL_b \ue^{\ui\alpha})}{\sin^2uL_b \ue^{\ui\alpha}}\\
&\sim \ue^{\ui\alpha} \Su \left(\frac{\cb - (1 - (uL_b\ue^{\ui\alpha})^2/2 + \ldots)}{uL_b\ue^{\ui\alpha}} + \frac{1 - \cb(1 - (uL_b\ue^{\ui\alpha})^2/2 + \ldots)}{uL_b\ue^{\ui\alpha}}\right)\\
&\sim u \ .
\end{align}
Therefore the integral \eqref{eq: zb} converges in the strip $0 < \Re(s) < 2$.

Combining the results,
\begin{equation}\label{eq: z1}
\zeta(s) = 2(\ue^{-\ui\pi s} + 1)\zeta_R(s)\Su\left(\frac{\pi}{L_b}\right)^{-s}
+ \ue^{\ui(\pi-\alpha)s}\frac{2\sin \pi s}{\pi}
\int_0^\infty u^{-s} \frac{\ud}{\ud u} \log f(ue^{\ui \alpha})\, \ud u \ ,
\end{equation}
for $0<\Re(s) < 2$.

To obtain an analytic continuation of $\zeta(s)$ also valid for $\Re(s) \leq 0$, we can split the integral at $u = 1$.  The restriction to $\Re(s)>0$ came from the behavior at infinity.
Let
\begin{equation}\label{eq: function f hat}
\hat{f}(u) = \Su \frac{\cb - \cos L_b\ue^{\ui \alpha}u}{\sin L_b\ue^{\ui \alpha} u} .
\end{equation}
Then
\begin{equation}
\hat{f}(u) = \ui\Su \frac{2\cb - (\ue^{\ui uL_b(\cos \alpha +\ui\sin \alpha)} + \ue^{-\ui uL_b(\cos \alpha+\ui \sin \alpha)})}{\ue^{\ui uL_b(\cos \alpha+\ui\sin \alpha)} - \ue^{-\ui uL_b(\cos \alpha +\ui \sin \alpha)}}
\sim \ui B \ ,
\end{equation}
in the limit $u \to \infty$.
Similarly
\begin{equation}
\frac{\ud}{\ud u}\hat{f}(u) = \ue^{\ui\alpha}\Su \frac{L_b(1 - \cb \cos uL_b\ue^{\ui\alpha} )}{\sin^2uL_b\ue^{\ui\alpha}}
\sim \ue^{-cu} ,
\end{equation}
for some $c>0$.
Therefore $\frac{\ud}{\ud u }\log \hat{f}(u) \sim \ue^{-cu}$ as $u \to \infty$.
Then, writing $\log f(u\ue^{\ui \alpha}) = \log u\ue^{\ui\alpha} + \log \hat{f}(u)$, one can expand the integral in (\ref{eq: z1})  to obtain the following theorem.

\begin{thm}\label{thm: rose zeta}
For the Dirac operator with zero mass on a rose graph with vertex conditions given by (\ref{eq: boundary condition 1}) and (\ref{eq: boundary condition 2}), the zeta function for $\Re(s) <2$ is given by,
\begin{align}\label{eq: z2}
\zeta(s) =&
\ue^{\ui(\pi-\alpha)s}\frac{2\sin \pi s}{\pi}
\left[ \int_0^1 u^{-s} \frac{\ud}{\ud u} \log u \ue^{\ui \alpha} \hat{f}(u) \, \ud u + \int_1^\infty u^{-s} \frac{\ud}{\ud u}\log \hat{f}(u)\, \ud u
+ \frac{1}{s} \right] \nn \\
& + 2(\ue^{-\ui\pi s} + 1)\zeta_R(s)\Su\left(\frac{\pi}{L_b}\right)^{-s}
\end{align}
where
\begin{equation}
\hat{f}(u) = \Su \frac{\cb - \cos L_b\ue^{\ui \alpha}u}{\sin L_b\ue^{\ui \alpha} u} .
\end{equation}
\end{thm}

\subsection{Spectral determinant of the rose graph}

As a corollary, it is straightforward to evaluate the spectral determinant from the zeta function.
The spectral determinant is formally the product of the eigenvalues,
\begin{equation}\label{eq: formal spectral det}
    {\det}' (\mathcal{D})=\prod_{j=-\infty}^\infty\,\!\!\!\!^{^\prime}\,\, k_j^2 \ ,
\end{equation}
where each eigenvalue $k_j$ has been squared in the product due to Kramers' degeneracy.  The representation of the zeta function in theorem \ref{thm: rose zeta} allows a regularized spectral determinant to be evaluated directly,
${\det}' (\mathcal{D})=\exp(-\zeta'(0))$.

Differentiating the pole contribution,
\begin{align}
 \zeta_p'(0) &= \ui\pi B - 2B\log(2\pi) + 2\Su \log \left(\frac{\pi}{L_b}\right) \label{eq:spec det 2} \ ,
\end{align}
where we used $\zeta_R(0)=-1/2$ and $\zeta'_R(0)=-\log(2\pi)/2$.

Notice that,
\begin{align}
\frac{\ud}{\ud s} \ue^{\ui(\pi-\alpha)s} \frac{2\sin \pi s}{\pi s}& =  2 \ue^{\ui(\pi-\alpha)s} \left( \ui(\pi-\alpha) \frac{\sin \pi s}{\pi s} + \frac{\pi s \cos \pi s - \sin \pi s}{\pi s^2} \right) \\
&\to2 \ui  (\pi -\alpha)
\end{align}
as $s\to 0$.  Hence,
\begin{align}
\zeta'(0)
&= \zeta_p'(0)+ \ui 2 (\pi -\alpha) \nn \\
&\quad +2\left( \log \left(\ue^{\ui\alpha} \hat{f}(1)\right) - \log \left( \Su \frac{\cb -1}{L_b} \right) +\log \left(\ui B \right) - \log \hat{f}(1) \right)\\
&= \ui\pi B - 2B\log(2\pi) + 2\Su \log \left(\frac{\pi}{L_b}\right) +2\pi \ui
- 2\log\left(\Su \frac{\cb - 1}{L_b} \right) + 2\log(\ui B) \ .
\end{align}
Therefore the spectral determinant is given by,
\begin{equation}\label{eq: spectral det rose}
 {\det}' (\mathcal{D})=
 \frac{ (-1)^{B+1}}{B^2} \left(\Su \frac{\cb - 1}{L_b} \right)^2 \prod_{b = 1}^B\left(2L_b\right)^{2} \ .
\end{equation}
Note that the spectral determinant is independent of $\alpha$, the angle of the branch cut of the logarithm, as expected.

\section{Zeta function of a general graph without mass}\label{sec: general zeta no mass}


The contour integral technique introduced in the case of a rose graph extends to any graph whose vertex matching conditions define a self-adjoint Dirac operator.
Let
\begin{align}\label{eq: f gen m=0}
    f(z)&=z^{4B-1}\det\left(\A + \B\begin{pmatrix}
\cot zL & -\csc zL\\
-\csc zL & \cot zL
\end{pmatrix}\right) \\
&= \frac{1}{z} \det\left(z\A + z\B\begin{pmatrix}
\cot zL & -\csc zL\\
-\csc zL & \cot zL
\end{pmatrix}\right) \ .
\end{align}
Notice that as $z\to 0$,
\begin{equation}\label{eq: f lim 0}
    \det\left(z\A + z\B\begin{pmatrix}
\cot zL & -\csc zL\\
-\csc zL & \cot zL
\end{pmatrix}\right) \to
 \det\left(\B\begin{pmatrix}
 L^{-1} & -L^{-1}\\
-L^{-1} & L^{-1}
\end{pmatrix}\right)=0 \ ,
\end{equation}
and the first non-zero term in its power series expansion about $z=0$ is therefore proportional to $z$.
Without loss of generality, we assume that $f(0)\neq 0$, as if $f(0)$ happens to be zero it can be made non-zero by a perturbation of the edge lengths.
Hence, $f$ has roots at the non-zero solutions of the secular equation \eqref{eq: general secular eqn m=0} and,
\begin{equation}\label{eq: asymptotics of f at zero}
f(z) \sim c_0 + c_Mz^M + O(z^{M+1})
\end{equation}
as $z \to 0$, with $c_0$, $c_M$ the first two non-zero coefficients in the power-series expansion which can be computed for any given graph with fixed vertex conditions.

For a general graph with zero mass, the spectral zeta function is then given by
\begin{equation}
\zeta(s) = \frac{1}{\ui\pi }\int _C z^{-s}\frac{\ud}{\ud z}\log f(z)\, \ud z
\end{equation}
where $C$ is the same contour that was used in the rose graph case, figure (\ref{fig:rose contour}).
Note that we also assume that the matrices $\A$ and $\B$ are independent of $k$; if $\A$ and $\B$ were $k$ dependent this could change the location of the zeros and poles of $f$.
Again, we develop the spectral zeta function by transforming the contour from $C$ to $C'$.  Then
\begin{equation}\label{eq: general graph three parts of zeta function}
\zeta(s) = \zeta_p(s) + \zeta_b(s) + \zeta_l(s)\ ,
\end{equation}
where $\zeta_p$ is the contribution from the poles of $f$, $\zeta_b$ is the integral around the branch cut, and $\zeta_l$ is the integral along the horizontal line (that will be sent to negative infinity in the imaginary coordinate).

The poles of $f$ are located at $z = n\pi/L_b, n \neq 0$, so as for the rose graph,
\begin{equation}\label{eq: general graph pole piece}
\zeta_p(s) = 2(\ue^{-\ui\pi s}+1)\zeta_R(s)\Su\left(\frac{\pi}{L_b}\right)^{-s} \ .
\end{equation}

On the horizontal line let $z = k + \ui t$.  In the limit $t\to -\infty$,
$\cot(zL) \to \ui$ and $\csc(zL) \to 0$, therefore
$f(z) \sim z^{4B-1}\det(\A +\ui\B)$.
This means that $\frac{\ud}{\ud z}\log f(z) \sim 1/z$ and consequently
$\zeta_l(s) \to 0$ for $\Re(s) > 0$.

For the integral around the branch cut, set $z =u \ue^{\ui \alpha}$ where the imaginary part of $z$ is positive, see figure \ref{fig:rose contour}. Then
\begin{equation}\label{eq: general case branch piece}
\zeta_b(s) = e^{\ui(\pi-\alpha)}\frac{2\sin\pi s}{\pi}\int_0^\infty u^{-s}\frac{\ud}{\ud u}\log f(ue^{\ui\alpha})\, \ud u \ .
\end{equation}
As $u \to \infty$,
\begin{equation}\label{eq: f hat lim infinity}
    \det\left(\A + \B\begin{pmatrix}
\cot uLe^{\ui\alpha} & -\csc uLe^{\ui\alpha}\\
-\csc uLe^{\ui\alpha} & \cot uLe^{\ui\alpha}
\end{pmatrix}\right) \to
 \det\left(\A - \ui\B\right),
\end{equation}
and hence $f(ue^{i\alpha}) \sim (ue^{\ui\alpha})^{4B-1}\det(\A - \ui\B)$. Consequently $\frac{\ud}{\ud u}\log f(ue^{\ui\alpha}) \sim 1/u$ and \eqref{eq: general case branch piece} converges at infinity provided that $\Re(s) > 0$.
From \eqref{eq: asymptotics of f at zero}
$\frac{\ud}{\ud u}\log f(ue^{\ui\alpha}) \sim  u^{M-1}$
as $u\to 0$.
Therefore \eqref{eq: general case branch piece} converges at zero for $\Re(s) < M$, and $\zeta_b(s)$ is defined in the strip $0 < \Re(s) < M$ where $M\geq 1$.

Combining the results,
\begin{equation}\label{eq: general zeta intermediate step}
\zeta(s) = 2(1+\ue^{-\ui\pi s} )\zeta_R(s)\sum_{b =1}^ B\left(\frac{\pi}{L_b}\right)^{-s} + \ue^{\ui(\pi - \alpha)s}\frac{2\sin{\pi s}}{\pi} \int_0^\infty u^{-s}\frac{\ud}{\ud u}\log f(u \ue^{\ui\alpha})\, \ud u
\end{equation}
for $0 < \Re(s) < M$.

To remove the restriction to $\Re(s) > 0$ we again split the integral at $u=1$.  Let
\begin{equation}
\hat{f}(u) =  \det\left(\A + \B\begin{pmatrix}
\cot u\ue^{\ui\alpha}L & -\csc u\ue^{\ui\alpha}L\\
-\csc u\ue^{\ui\alpha}L & \cot u\ue^{\ui\alpha}L
\end{pmatrix}\right).
\end{equation}
Since $\Im \ue^{\ui\alpha} > 0$,
\begin{equation}
\hat{f}(u) \sim \det\left(\A - \ui\B\begin{pmatrix}
\ue^{2\ui L\ue^{\ui\alpha}u} + 1 & -2\ue^{\ui L\ue^{\ui\alpha}u}\\
-2\ue^{\ui L\ue^{\ui\alpha}u} & \ue^{2\ui L\ue^{\ui\alpha}u} + 1
\end{pmatrix}\right)
\end{equation}
as $u \to \infty$.  Consequently $\hat{f}'/\hat{f}$ decays exponentially as $u\to \infty$.  Then splitting the integral at $u=1$ and using $\log f(u\ue^{\ui \alpha}) = \log(\hat{f}(u)) + (4B-1)(\ui \alpha + \log u)$ to develop the integral from $1$ to $\infty$, we obtain the following theorem.

\begin{thm}\label{thm: general massless zeta}
For the Dirac operator with zero mass on a graph with local vertex matching conditions defined by an energy independent pair of matrices $\A$ and $\B$ with $\A\B^\dagger = \B\A^\dagger$, the zeta function is given by,
\begin{align}\label{eq: general graph zeta function}
\zeta(s) = &\ue^{\ui(\pi - \alpha)s}\frac{2\sin{\pi s}}{\pi} \left[\int_0^1 u^{-s}\frac{\ud}{\ud u}\log \left((u\ue^{\ui\alpha})^{4B-1}\hat{f}(u)\right)\ud u + \int_1^\infty u^{-s}\frac{\ud}{\ud u}\log\hat{f}(u)\ud u \right]\nonumber\\
&+\ue^{\ui(\pi - \alpha)s}\frac{2(4B-1)\sin{\pi s}}{\pi s} + 2(1+\ue^{-\ui\pi s})\zeta_R(s)\Su \left(\frac{\pi}{L_b}\right)^{-s} \ ,
\end{align}
where $\Re(s) < M$ for some $M>1$ and
\begin{equation}\label{eq: general graph f hat function}
\hat{f}(u) =  \det\left(\A + \B\begin{pmatrix}
\cot u\ue^{\ui\alpha}L & -\csc u\ue^{\ui\alpha}L\\
-\csc u\ue^{\ui\alpha}L & \cot u\ue^{\ui\alpha}L
\end{pmatrix}\right).
\end{equation}
\end{thm}

\subsection{Spectral determinant with zero mass}

The regularized spectral determinant is
${\det}' (\mathcal{D})=\exp(-\zeta'(0))$.
Using the analogy with the rose graph case, which has the same pole contribution,
\begin{align}
\zeta'(0)&= \zeta_p'(0)+2\left(\log \left(f(\ue^{\ui\alpha})\right) - \log \left(\lim_{u \to 0} f(u\ue^{\ui \alpha})\right)+\log \left(\lim_{u \to \infty} \hat{f}(u)\right) - \log\left(\hat{f}(1)\right)\right) \nn \\
& \quad+2\ui(4B-1)(\pi-\alpha) \nn\\
&= \zeta_p'(0)+2\ui(4B-1)\alpha - 2\log c_0 + 2\log(\det(\A - \ui\B)) +2\ui(4B-1)(\pi-\alpha) \nn\\
&= \ui \pi B - 2B \log (2\pi) +2 \sum_{b=1}^B \log \left( \frac{\pi}{L_b} \right) -2\log c_0 \nn\\
& \quad + 2\log(\det(\A - \ui\B)) +2\ui(4B-1)\pi \label{eq: general spec det 1}
\end{align}
where we used $f(e^{\ui\alpha}) = \ue^{\ui(4B-1)\alpha}\hat{f}(1)$ and \eqref{eq: f hat lim infinity}.  The constant $c_0$ comes from the power series expansion of $f$ about zero; see \eqref{eq: asymptotics of f at zero}.
Therefore the spectral determinant is,
\begin{equation}\label{eq: spectral det no mass}
 {\det}' (\mathcal{D})=
 \frac{ {c_0}^2  (-1)^{B}}{(\det(\A - \ui\B))^2} \prod_{b = 1}^B\left(2L_b\right)^{2} \  .
\end{equation}

\section{Zeta function of a general graph with mass}\label{sec: general zeta mass}

To derive the zeta function with non-zero mass we start from the two secular equations
\eqref{eq: general secular eqn} and  \eqref{eq: general secular eqn -ve E} whose roots correspond to positive and negative energy solutions respectively.
Let
\begin{align}\label{eq: f gen mass}
f(z) & = \det\left(\A + \gamma(z)\B\begin{pmatrix}
\cot zL & -\csc zL\\
-\csc zL & \cot zL
\end{pmatrix}\right) \ , \\
\hat{f}(t) & = \det\left(\A + \hat{\gamma}(t)\B\begin{pmatrix}
\coth tL & -\csch tL\\
-\csch tL & \coth tL
\end{pmatrix}\right) \ ,
\end{align}
where
\begin{align}
\gamma(z) &= \frac{\sqrt{z^2 + m^2} - m}{z} \ , \\
\hat{\gamma}(t) &= \frac{\sqrt{t^2 - m^2} +\ui m}{t} \ .
\end{align}
So $f(\ui t)=\hat{f}(t)$ and the positive energy secular equation reads $f(k)=0$ for $k>0$. Similarly we set,
\begin{align}\label{eq: f gen mass}
g(z) & = \det\left(\gamma(z)\A - \B\begin{pmatrix}
\cot zL & -\csc zL\\
-\csc zL & \cot zL
\end{pmatrix}\right) \ , \\
\hat{g}(t) & = \det\left(\hat{\gamma}(t)\A - \B\begin{pmatrix}
\coth tL & -\csch tL\\
-\csch tL & \coth tL
\end{pmatrix}\right) \ ,
\end{align}
so the negative energy secular equation reads $g(k)=0$ for $k>0$.

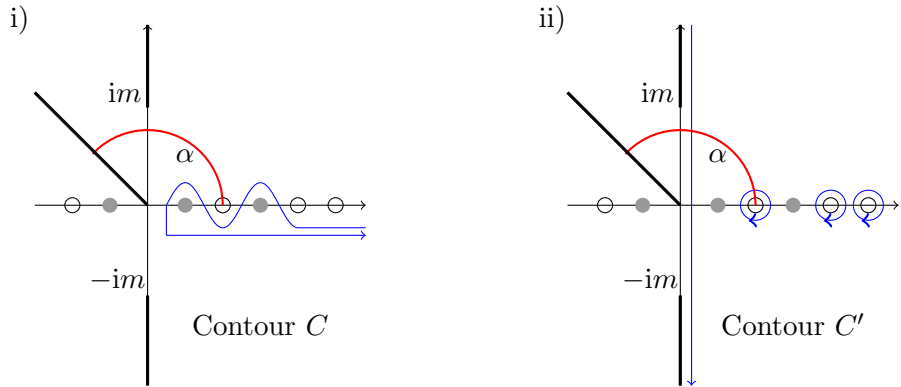
\begin{figure}[tbhp!]

\begin{center}
\begin{tikzpicture}
  \draw[->] (-1.5,0) -- (2.9,0);
  \draw[->] (0,-2.4) -- (0,2.4);
  \draw (-1,0) circle (.1cm);
  \draw (1,0) circle (.1cm);
  \draw (2,0) circle (.1cm);
  \draw (2.5,0) circle (.1cm);
  \fill[black!40!white] (-.5,0) circle (.1cm);
  \fill[black!40!white] (.5,0) circle (.1cm);
  \fill[black!40!white] (1.5,0) circle (.1cm);
  \draw[blue] (.25,0) sin (.5,.3);
  \draw[blue] (.5,.3) cos (.75,0);
  \draw[blue] (.75,0) sin (1,-.3);
  \draw[blue] (1,-.3) cos (1.25,0);
  \draw[blue] (1.25,0) sin (1.5,.3);
  \draw[blue] (1.5,.3) cos (1.75,0);
  \draw[blue] (1.75,0) sin (2,-.3);
  \draw[blue] (2,-.3) -- (2.9,-.3);
  \draw[blue] (.25,0) -- (.25,-.4);
  \draw[->, blue](0.25,-.4) -- (2.9,-.4);
  \draw[very thick] (0,1.3) -- (0,2.4);
  \draw[very thick] (0,-1.2) -- (0,-2.4);
  \draw[red,thick,domain=0:137] plot ({cos(\x)}, {sin(\x)});
  \draw[very thick] (-1.5,1.5) -- (0,0);
  \node[label={\small $\alpha$}] at (.5,.3){};
  \node[label={\small Contour $C$}] (a) at (1.5,-2){};
  \node[label ={\small $\ui m$}] at (-.3,1.1){};
  \node[label ={\small $-\ui m$}] at (-.4,-1.4){};
  \node[label={\small i)}] (a) at (-1.7,2){};
\end{tikzpicture}
\hspace{.75in}
\begin{tikzpicture}[decoration={markings,
mark=at position -0.3cm with {\arrow[line width=1pt]{<}},
}
]
  \draw[->] (-1.5,0) -- (2.9,0);
  \draw[->] (0,-2.4) -- (0,2.4);
  \draw (-1,0) circle (.1cm);
  \draw (1,0) circle (.1cm);
  \draw (2,0) circle (.1cm);
  \draw (2.5,0) circle (.1cm);
  \fill[black!40!white] (-.5,0) circle (.1cm);
  \fill[black!40!white] (.5,0) circle (.1cm);
  \fill[black!40!white] (1.5,0) circle (.1cm);
  \draw[blue,postaction=decorate] (1,0) circle (.2cm);
  \draw[blue,postaction=decorate] (2,0) circle (.2cm);
  \draw[blue,postaction=decorate] (2.5,0) circle (.2cm);
  \draw[->, blue] (.15,2.4) -- (.15,-2.4);
  \draw[very thick] (0,1.3) -- (0,2.4);
  \draw[very thick] (0,-1.2) -- (0,-2.4);
  \draw[red,thick,domain=0:137] plot ({cos(\x)}, {sin(\x)});
  \draw[very thick] (-1.5,1.5) -- (0,0);
  \node[label={\small $\alpha$}] at (.5,.3){};
  \node[label={\small Contour $C'$}] at (1.5,-2){};
  \node[label ={\small $\ui m$}] at (-.3,1.1){};
  \node[label ={\small $-\ui m$}] at (-.4,-1.4){};
  \node[label={\small ii)}] (a) at (-1.7,2){};
\end{tikzpicture}
\end{center}

\caption{\label{fig: general contour}Contours $C$ and $C'$. We again locate the branch cut of the logarithm at an angle $\alpha$ and branch cuts of $(z+\ui m)^{-s/2}$ and $(z-\ui m)^{-s/2}$ are located on the imaginary axis.   The zeros and poles of $f$ (or $g$) are shown with filled and empty circles respectively.}
\end{figure}

The contribution of the positive part of the spectrum to the spectral zeta function is
\begin{equation}
\zeta^+(s) = \frac{1}{\ui\pi }\int _C (z^2 + m^2)^{-s/2}\frac{\ud}{\ud z}\log f(z)\, \ud z \ ,
\end{equation}
where the contour $C$, shown in figure \ref{fig: general contour}, is chosen to enclose the zeros of $f$ while avoiding poles.
We locate the branch cut of the logarithm at an angle $\alpha$ and branch cuts of
$(z+\ui m)^{-s/2}$ and $(z-\ui m)^{-s/2}$ on the imaginary axis.
Transforming the contour from $C$ to $C'$, see figure \ref{fig: general contour} (ii),
\begin{equation}\label{eq: general graph mass three parts of zeta function}
\zeta^+(s) = \zeta_p^+(s) + \zeta^+_b(s) \ ,
\end{equation}
where $\zeta^+_p$ is the contribution from the poles of $f$ and $\zeta^+_b$ is given by the integral along the imaginary axis.

The poles of $f$ occur at $z = n\pi/L_b, n \neq 0$ so
\begin{align}\label{eq: gen poles}
\zeta^+_p(s) &= 2\Su \sum_{n=1}^\infty \left(\left(\frac{n\pi}{L_b}\right)^2 + m^2\right)^{-s/2} \nn\\
  & = 2 \Su \left( \frac \pi {L_b} \right)^{-s} E \left( \frac s 2 , \left( \frac{ mL_b} \pi \right)^2 \right),
\end{align}
where $E (\alpha , c)$ is an Epstein type zeta function \cite{epst03-56-615,epst07-63-205,eliz89-30-1133,kirs94-35-459}
\begin{eqnarray}
E (\alpha , c) = \sum_{n=1}^\infty (n^2+c)^{-\alpha} . \label{epst1}
\end{eqnarray}
For the imaginary axis integral let $z=\ui t+\epsilon$ for some sufficiently small $\epsilon >0$.
\begin{align}
\zeta^+_b(s) &= \frac{1}{\ui \pi} \int_\infty^{-\infty} \left( (\ui t +\epsilon )^2+m^2\right)^{-\frac{s}{2}} \frac{\ud}{\ud t} \log \hat{f}(t) \, \ud t  \nn \\
& = - \frac{1}{\ui \pi} \int_0^\infty \left( (\ui t +\epsilon )^2+m^2\right)^{-\frac{s}{2}} \frac{\ud}{\ud t} \log \hat{f}(t) \, \ud t+ \frac{1}{\ui \pi} \int_0^\infty \left( (\ui t -\epsilon )^2+m^2\right)^{-\frac{s}{2}} \frac{\ud}{\ud t} \log \hat{f}(t) \, \ud t \nn  \\
& = \frac{2}{\pi} \sin\left( \frac{\pi s}{2}\right) \int_m^\infty  (t^2-m^2)^{-\frac{s}{2}} \frac{\ud}{\ud t} \log \hat{f}(t) \, \ud t \ ,
\end{align}
where we used $\hat{f}(t)=\hat{f}(-t)$ and we take the limit $\epsilon \to 0$ in the last step. We note that

\begin{align}
    \hat{\gamma}'(t)&= (t^2-m^2)^{-\frac{1}{2}} - \left( \frac{(t^2-m^2)^{\frac{1}{2}} +\ui m}{t^2}\right)\\
    &= \frac{m^2-\ui m (t^2-m^2)^{\frac{1}{2}}}{t^2(t^2-m^2)^\frac{1}{2}} \ . \label{eq: gamma'}
\end{align}
Hence as $t\to m^+$, the integrand behaves as $(t^2-m^2)^{-\frac{(s+1)}{2}}$ and the integral converges at $t=m$ for $\Re s < 1$.

As $t\to \infty$
up to exponentially damped terms,
\begin{align}
\hat{f}(t) &\sim \det (\A+\hat{\gamma}(t) \B) \\
& \sim \det (\A+\B) +c_{1} t^{-1}+\dots+c_{4B} t^{-4B} \ .
\end{align}
So $ \hat{f}'/\hat{f}  \sim t^{-2}$ and the integral converges at infinity for $\Re s > -1$.

The contribution of the negative energy eigenvalues is evaluated similarly, namely
\begin{equation}
\zeta^-(s) = \frac{(-1)^{-s}}{\ui\pi }\int _C (z^2 + m^2)^{-s/2}\frac{\ud}{\ud z}\log g(z)\, \ud z \ ,
\end{equation}
where $C$ is now chosen to enclose the zeros and avoid the poles of $g$.  Applying the same contour transformation,
\begin{equation}\label{eq: general graph mass negative zeta function}
\zeta^-(s) = \zeta_p^-(s) + \zeta^-_b(s)
\end{equation}
with the pole contribution,
\begin{equation}
\zeta_p^-(s) = (-1)^{-s} \zeta^+_p(s) \  ,
\end{equation}
and
\begin{align}
\zeta^-_b(s) &= \frac{2(-1)^{-s}}{\pi} \sin\left( \frac{\pi s}{2}\right) \int_m^\infty  (t^2-m^2)^{-\frac{s}{2}} \frac{\ud}{\ud t} \log \hat{g}(t) \, \ud t \  ,
\end{align}
which is convergent again in the strip $-1<\Re s < 1$. Combining the results, we obtain the following theorem.

\begin{thm}\label{thm: general zeta with mass}
For the Dirac operator with non-zero mass on a graph with local vertex matching conditions defined by a pair of $k$ independent matrices $\A$ and $\B$ with $\A\B^\dagger = \B\A^\dagger$, the zeta function in the strip
$-1<\Re s < 1$ is given by
\begin{align*} 
\zeta(s) =&\frac{2}{\pi} \sin\left( \frac{\pi s}{2}\right)\left( \int_m^\infty  (t^2-m^2)^{-\frac{s}{2}} \frac{\ud}{\ud t} \log \hat{f}(t) \, \ud t
+ (-1)^{-s} \int_m^\infty  (t^2-m^2)^{-\frac{s}{2}} \frac{\ud}{\ud t} \log \hat{g}(t) \, \ud t
\nn \right) \\
&+2(1+(-1)^{-s}) \Su \left( \frac{\pi}{L_b}\right)^{-s}  E \left( \frac s 2 , \left( \frac{ mL_b} \pi \right)^2 \right)       \  .
\end{align*}
\end{thm}

\subsection{Spectral determinant with mass}

Again the regularized spectral determinant is ${\det}' (\mathcal{D})=\exp(-\zeta'(0))$.  Differentiating the pole contribution,
\begin{align}
\zeta_p^\prime(0)&=\Su  \left[4\left\{ \frac 1 2 E' \left( 0, \left( \frac{mL_b} \pi \right)^2 \right)
+ \log \left(\frac{L_b}{\pi} \right) E \left( 0,\left( \frac{mL_b}{\pi} \right)^2 \right)\right\} \right.\nn\\
& \quad \left.+2\pi \ui E \left( 0, \left(\frac{mL_b}{\pi} \right)^2\right) \right]\ .
\end{align}
Using the analytical continuation of the Epstein zeta function \cite{eliz89-30-1133},
\begin{eqnarray}
E (\alpha , c) &=& - \frac 1 {2 c^\alpha} + \frac{ \sqrt \pi} {2 c^{\alpha - 1/2} \Gamma (\alpha )} \times \nn\\
& & \left\{ \Gamma \left( \alpha - \frac 1 2 \right) + 4 \sum_{\ell =1} ^\infty \frac 1 {(\pi \ell \sqrt c)^{1/2 - \alpha}} K_{1/2 - \alpha }
(2\pi \ell \sqrt c)\right\} , \label{epst2}
\end{eqnarray}
with $K_u (z)$ the Kelvin functions \cite{grad65b}, it is easily verified that
\begin{eqnarray}
E ( 0,c) = - \frac 1 2 \label{epst3}
\end{eqnarray}
and
\begin{eqnarray}
E' (0,c) = \frac 1 2 \log c- \pi \sqrt c - \log \left( 1 - e^{-2\pi \sqrt c} \right). \label{epst4}
\end{eqnarray}
From here we find
\begin{eqnarray}
\zeta _p ' (0) = 2 \sum_{b=1}^B \left[ \log \left( \frac m {1-e^{-2mL_b}}\right) -mL_b \right] - i \pi B. \label{polenew}
\end{eqnarray}
Differentiating the integral terms,
\begin{align}
\zeta_b'(0) & = \lim_{t\to \infty}\log \hat{f}(t) -\log  \hat{f}(m)  +\lim_{t\to \infty} \log \hat{g} (t) -\log  \hat{g}(m) \\
&=  \log \det\left(\A + \B\right) + \log \det\left(\A - \B\right)  -2\log \det\left(\A + \ui \B\begin{pmatrix}
\coth mL & -\csch \, mL\\
-\csch \, mL & \coth mL
\end{pmatrix}\right)  \ .
\end{align}
Hence the spectral determinant is, after elementary simplifications,
\begin{align}\label{eq: spectral det with mass}
 &{\det}' (\mathcal{D})=
 \left( \det\left(\A + \ui \B\begin{pmatrix}
\coth mL & -\csch \, mL\\
-\csch \, mL & \coth mL
\end{pmatrix}\right) \right)^2(\det(\A +\B))^{-1}(\det(\A -\B))^{-1} \nn \\
&\quad \quad \quad \quad \quad \times (-1)^B\prod_{b = 1}^B \left[ \frac{ 2\sinh (mL_b)} m \right]^2 \  .
\end{align}

It is important to note that, the representation of the spectral zeta function in theorem \ref{thm: general zeta with mass} for $m>0$ is not valid in the limit $m\to 0$ where the integrals diverge.  Consequently one cannot take $m \to 0$ in the spectral determinant \eqref{eq: spectral det with mass} and the zero mass case must be obtained independently, as was done in the previous section.  Despite this, formulas for the spectral determinant with and without mass, equations \eqref{eq: spectral det with mass} and \eqref{eq: spectral det no mass}  respectively, show a number of common features.  

\section{Conclusions}

We have constructed the spectral zeta functions of the Dirac operator on finite quantum graphs.
The contour integral technique we employ allows one to analyze general graphs with general, local, energy-independent, vertex conditions in a single calculation.  Results for individual graphs are straightforward to extract from the general case given a pair of matrices $\A$ and $\B$ specifying the vertex conditions.  The approach combines results for the secular equations of the Dirac operator with the argument principle and allows us to obtain an integral representation of the zeta function valid in a strip of the complex plane, theorems \ref{thm: general massless zeta} and \ref{thm: general zeta with mass}.  Analytic continuation to any region of the plane can be obtained from the asymptotic behavior of the secular equation.
As a special case, we see that the zeta function of the Dirac rose graph has a particularly simple form, theorem \ref{thm: rose zeta}.  In each case, as a straightforward corollary, we obtained the regularized spectral
determinant, equations \eqref{eq: spectral det rose}, \eqref{eq: spectral det no mass} and \eqref{eq: spectral det with mass}.

\begin{acknowledgments}
The authors would like to thank Rachel Wilkerson for helpful suggestions.  This work was partially supported by a grant from the Simons Foundation (354583 to Jon Harrison).
\end{acknowledgments}

\appendix
\section{Derivation of secular equation for a Dirac rose graph}\label{sec: secular equation derivation}
As the test case of the Dirac operator on a rose graph, shown in figure \ref{fig: rose graph}, is much less well known than the corresponding case of the Laplace operator on a star graph with Neumann-like matching conditions, we include the derivation of the secular equation \cite{HarWin_JPA12}.  Importantly the secular equation can be expressed without using matrices despite the spinor valued nature of the wavefunction on the bonds and the nontrivial spin dynamics at the vertex.

We split the four component wavefunctions on the bonds of the rose into pairs of two component functions,
\begin{equation}\label{eq: v and w}
v^b(x_b) = \begin{pmatrix} \psi^b_1(x_b)\\\psi^b_2(x_b)\end{pmatrix} \hspace{1cm}\mbox{and}\hspace{1cm} w^b(x_b) = \begin{pmatrix}-\psi^b_4(x_b)\\\psi^b_3(x_b)\end{pmatrix}.
\end{equation}
Then the rose graph is given by matching conditions at the vertex such that firstly,
\begin{equation}\label{eq: continuity condition}
u^b_o v^b(0) = u^b_t v^b(L_b) = \xi \hspace{1cm}\mbox{for all $b$,}
\end{equation}
where $\xi$ is used not as a fixed vector but rather as a placeholder for the common value at the vertex.   Secondly,
\begin{equation}\label{eq: derivative boundary condition}
\hspace{-2cm}\sum_{b=1}^B u^b_o w^b(0) = \sum_{b=1}^B u^b_t w^b(L_b) \ .
\end{equation}
The $u^b_{o/t}$ are $\SU(2)$ matrices with $o,t$ standing for the element of $\SU(2)$ associated with the origin and terminus of the bond $b$ respectively.  Such a vertex condition defines a self-adjoint realization of the Dirac operator on the graph\cite{BH03}.  The definition is analogous to the definition of Neumann-like vertex conditions for the Laplace operator.  Applying such vertex conditions to a rose graph rather than a star graph is required in order to produce nontrivial spin dynamics.

Using the plane-wave solution \eqref{eq: eigenspinors} for zero mass, \eqref{eq: continuity condition} becomes
\begin{equation}\label{eq: continuity condition evaluated}
u^b_o(\bmu^b + \bhmu^b) = u^b_t(\bmu^b \ue^{\ui kL_b} + \bhmu^b \ue^{-\ui kL_b}) = \xi,
\end{equation}
where
\begin{equation}
\bmu^b = \begin{pmatrix} \mu_\alpha^b\\\mu^b_\beta\end{pmatrix} \hspace{1cm}\mbox{and}\hspace{1cm} \bhmu^b = \begin{pmatrix}\hat{\mu}_\alpha^b\\\hat{\mu}^b_\beta\end{pmatrix} \ .
\end{equation}
Then
\begin{equation}\label{eq: continuity condition rearranged}
(u^b_o - u^b_t\ue^{\ui kL_b})\bmu^b = -(u^b_o - u^b_t\ue^{-\ui kL_b})\bhmu^b \ ,
\end{equation}
and
\begin{equation}\label{eq: mu plus mu hat}
\bmu^b + \bhmu^b = (u^b_o)^{-1}\xi \ .
\end{equation}
Eliminating $\hat{\mu}^b$
\begin{equation}\label{eq: formula for mu}
\bmu^b = \frac{1}{2\ui \sin kL_b}(u^b_t)^{-1}(u^b_o - u^b_t\ue^{-\ui kL_b})(u^b_o)^{-1}\xi
\end{equation}
 assuming $\sin kL_b \neq 0$.  The second part of the vertex condition \eqref{eq: derivative boundary condition} reads
\begin{equation}
\ui\gamma(k)\sum_{b=1}^B u^b_o(-\bmu^b + \bhmu^b) = \ui\gamma(k)\sum_{b=1}^B u^b_t(-\bmu^b\ue^{\ui kL_b} + \bhmu^b\ue^{-\ui kL_b}),
\end{equation}
which simplifies to
\begin{equation}\label{eq: simplified summation}
\sum_{b=1}^B \left(u^b_t\ue^{\ui kL_b} - u^b_o(b)\right)\bmu^b = \mathbf{0}
\end{equation}
using equation \eqref{eq: continuity condition rearranged}. Substituting \eqref{eq: formula for mu} into \eqref{eq: simplified summation},
\begin{align}
\sum_{b=1}^B \frac{1}{\sin kL_b} \left(u^b_t\ue^{\ui kL_b} - u^b_o\right)(u^b_t)^{-1}\left(u^b_o - u^b_t\ue^{-\ui kL_b}\right)(u^b_o)^{-1}\xi &= \mathbf{0}\nonumber\\
\sum_{b=1}^B \frac{1}{\sin kL_b}\left(\ue^{\ui kL_b}I_2 - u^b_o(u^b_t)^{-1}\right)\left(I_2 - u^b_t(u^b_o)^{-1}\ue^{-\ui kL_b}\right)\xi &= \mathbf{0}\nonumber\\
\sum_{b=1}^B \frac{1}{\sin kL_b}\left(2\cos (kL_b) \, \UI_2 - u^b_t(u^b_o)^{-1} - u^b_o(u^b_t)^{-1}\right)\xi &= \mathbf{0}.\label{eq: condition for eigenvalue}
\end{align}
For each bond we may define an angle $\theta_b\in [0,\pi]$ via,
\begin{equation}\label{eq: trace of matrices}
 u^b_t(u^b_o)^{-1} + u^b_o(u^b_t)^{-1}=\mbox{tr}\left(u^b_o(u^b_t)^{-1}\right) \UI = 2\cos \theta_b \UI \ .
\end{equation}
Then  $E(k) = \sqrt{k^2 + m^2}$ is an energy eigenvalue of the rose graph if and only if $k$ is a solution of
\begin{equation}
\sum_{b=1}^B \frac{\cos \theta_b - \cos kL_b}{\sin kL_b} = 0 \ .
\end{equation}
This is the secular equation from which we derived our results for the rose graph.


\bibliographystyle{abbrv}
\bibliography{dirac_bib}
\end{document}